\documentclass[prd,twocolumn,amsmath,nofootinbib]{revtex4-1}

\usepackage{amssymb,amsmath}
\usepackage{color}
\usepackage{graphicx}
\usepackage[normalem]{ulem}
\usepackage{graphics}
\usepackage{soul}
\usepackage[dvipsnames]{xcolor}
\usepackage{mathtools}
\usepackage[normalem]{ulem}
\usepackage{subfigure}
\usepackage{bm}
\usepackage{physics}
\usepackage{hyperref}

\begin{document}

\title{An angular momentum based graviton detector}

\author{J. P. M. Pitelli}
\email[]{pitelli@unicamp.br}
\affiliation{Departamento de Matem\'atica Aplicada, Universidade Estadual de Campinas,
13083-859 Campinas, S\~ao Paulo, Brazil}%

\author{T. Rick Perche}
\email[]{trickperche@perimeterinstitute.ca}

\affiliation{Perimeter Institute for Theoretical Physics, Waterloo, Ontario, N2L 2Y5, Canada}
\affiliation{Department of Applied Mathematics, University of Waterloo, Waterloo, Ontario, N2L 3G1, Canada}

\begin{abstract}

We show that gravitons with energy $E<\Omega$, where $\Omega$ is the energy gap a localized non-relativistic system, can be detected by finite-time interactions with a detector. Our detector is based on a quadrupole moment interaction between the hydrogen atom and the gravitational field in the linearized approximation. In this model, the external agent responsible for switching the interaction on an off inputs energy into the system, which creates a non-zero excitation probability even when the field is in the vacuum state. However, when the gravitational field is in a one-particle state with angular momentum, we obtain excitations due to the field's particle content. These detector excitations are then associated with the detection of gravitons. We also discuss a possible physical realization of our model where the  electromagnetic field plays the role of the external agent. 
\end{abstract} 


\maketitle

\section{\label{sec:intro}Introduction}

In the low energy regime,  the gravitational content can be viewed as a perturbation of the metric around the Minkowski background. In this linearized theory of gravity, the perturbation satisfies a linear equation of motion and can be quantized by canonical quantization methods~\cite{weinberg,weinbergBook}. The excitations of the  resulting spin-2 quantum field are called gravitons. Even though we believe the gravitons mediate the gravitational force, they had never been experimentally detected, the reason for this being the extraordinary weakness of the gravitational interaction. 

Following Dyson's proposition that gravitons cannot be measured by any conceivable apparatus~\cite{dysonReview}, the detection of a single  graviton by a hydrogen atom was considered in details in Refs.~\cite{gravDetec,Rothman}. The graviton emission and absorption by a hydrogen atom, as well as the ionization cross section were calculated. It was found that, even the detection  being in principle possible, it is seems to be highly unlikely using this technology since the typical  cross-sections are of the order of the  Planck length squared. In this paper we do not intend to solve this issue, but to shed light on a new way of detecting soft gravitons (by this we mean gravitons with energies below a certain threshold $\Omega$) still using the hydrogen atom as our detector. We should emphasize that our method can in principle be used to measure any tensorial field via dipole or quadrupole interactions, in particular photons. However, since polarized photons will play the role of the external agent in our physical realization (see Sec.~\ref{implementing}), we illustrate our model by probing the gravitational field.

The definition of particles in a general curved spacetime has many ambiguities and subtleties~\cite{birrell_davies}. {In fact, even in  Minkowski spacetime, the particle concept becomes ambiguous under a general change of reference frame~\cite{fullingUnruhEffect}.} One approach to this issue  was introduced by Unruh in~\cite{Unruh1976}, where he introduced the concept of particle detectors for the first time. His definition of a particle detector consisted of a non-relativistic quantum system confined in a box. This system would then interact with a quantum field and the excitations produced by this interaction would be regarded as particles. Unruh's solution to the ambiguity of the definition of particles can be summarized in his famous statement that ``a particle is what a particle detector detects''.  

Few years after Unruh's proposal of a particle detector, DeWitt simplified the model to a two-level system defined along a trajectory~\cite{DeWitt}, defining the model that would later be called the Unruh-DeWitt (UDW) detector. In essence, this particle detector model is a two-level system that couples to a scalar field via a monopole moment interaction. Although simple, the UDW model has been shown to capture the fundamental features of the interaction of non-relativistic systems with quantum fields. In fact, it has been shown to model some features of the light-matter interaction better that the typical models used in quantum optics~\cite{eduardoOld,eduardo,Nicho1,richard}. The UDW model can also be adapted to describe the interaction of nucleons with the three flavour neutrino fields, yielding results that match the phenomenology of this process~\cite{neutrinos,antiparticles}. Moreover, the UDW model and more general instances of particle detectors~\cite{mineInPreparation} can be argued to be the most adequate tools when approaching reference frame dependent phenomena in quantum field theory, such as the Unruh and Hawking effects~\cite{HawkingGibons,Unruh-Wald,Takagi,bhDetectorsBTZ,bhDetectors,bhDetectorsAdS}, where observers in a given state of motion perceive a thermal bath of particles. Another important application of particle detector models is to probe informational aspects of quantum field theory, quantifying the entanglement within quantum fields~\cite{Valentini1991,Reznik2003,Reznik1,Pozas-Kerstjens:2015,Pozas2016,Nick,topology} and their informational capacity~\cite{Katja,Simidzija_2020}. 

Although very successful, the UDW model is far from being the most general particle detector model. In fact, any localized non-relativistic quantum system that couples to a quantum field theory defines a particle detector. For instance, the hydrogen atom is one of such systems. It has a dipole and a quadrupole moment and it interacts with both the electromagnetic~\cite{Pozas2016} and gravitational~\cite{remi} fields. When coupled to the quantized gravitational field, it can viewed as a graviton detector, exchanging energy and angular momentum with the gravitational field. In the adiabatic limit, where the interaction is effectively turned on for arbitrary long times, the transient effects are washed out so that energy and angular momentum are conserved in the process of absorption or emission of gravitons~\cite{gravDetec}. However, for finite time interactions, the effect of switching the interaction on and off requires an external agent to input energy into the system, and energy/angular momentum are not generally conserved. Nevertheless, if the interaction of the atom with the quantum field is switched in a way that preserves the spherical symmetry of the system, angular momentum must still be a conserved quantity. We explore this idea to propose a way of detecting particles that relies on the exchange of angular momentum rather than energy. 

This paper is organized as follows. In Section \ref{sec:linQG} we review the formalism of linearized quantum gravity and set the conventions that will be adopted throughout the manuscript. In Section \ref{sec:gMatter} we detail the interaction of a hydrogen-like atom with a weak gravitational field. Section \ref{sec:main} is the main section of the manuscript, where we present the resulting probability of detection of gravitons through angular momentum exchanges. In Section \ref{implementing} we discuss a physical implementation of our setup, focusing on the implementation of a finite time interaction. Our conclusions can be found in Section \ref{sec:conclusion}.

{\color{black} Throughout this manuscript we work with Planck units, so that physical quantities are expressed in terms of Planck length and mass.}

\section{Linearized Quantum Gravity}\label{sec:linQG}

Einstein's theory of general relativity is among the most successful theories of physics. It introduces dynamics to the background spacetime, which is curved by the presence of matter according to Einstein's equations,
\begin{equation}
    G_{\mu\nu}= \frac{8 \pi}{\color{black} m_p^2} T_{\mu\nu},
\end{equation}
{\color{black} where $m_p$ denotes the  Planck mass.} However, despite its success, the fact that Einstein's equations depend on a classical description for matter and are non-linear on the metric make it hard to connect general relativity with quantum theory.

One of the regimes in which we can formulate a quantum theory for the spacetime metric is in the linearized regime. That is, we consider a metric perturbation on a fixed background and only take into account linear terms in the perturbation. The most common setup to apply this technique is in Minkowski spacetime, where we consider the metric
\begin{equation}
    g_{\mu\nu} = \eta_{\mu\nu} + h_{\mu\nu},
\end{equation}
where $\eta_{\mu\nu} = \text{diag}(-1,1,1,1)$ is the Minkowski metric and $h_{\mu\nu}$ is the perturbation. The equations of motion for the field $h_{\mu\nu}$ will then be linear to leading order, by construction, and in the absence of matter ($T_{\mu\nu} = 0$). In inertial coordinates they read
\begin{equation}\label{linEin}
    2\partial_{\sigma} \partial_{(\mu} h^{\sigma}{}_{\nu)}-\partial_{\mu} \partial_{\nu} h-\square h_{\mu \nu}+\eta_{\mu \nu}( \square h- \partial_{\rho} \partial_{\lambda} h^{\rho \lambda}) = 0,
\end{equation}
where $h = \eta_{\mu\nu}h^{\mu\nu}$ and $\square = \partial_\mu\partial^\mu$. Equation \eqref{linEin} is also invariant under infinitesimal coordinate transformations, which is associated with the gauge symmetry of general relativity. We will use this gauge freedom and work in the traceless transverse gauge associated with a given inertial observer with four-velocity $u^\mu$~\cite{Wald1,weber}, so that $h = 0$, $\partial_\mu h^{\mu\nu} = 0$ and $u^\mu h_{\mu\nu} = 0$.

From the linear equations of motion for the metric perturbation, it is possible to write a consistent free quantum  field theory for $h_{\mu\nu}$. In the traceless transverse gauge, it is then possible to write the expansion of the quantized field $\hat{h}_{\mu\nu}$ as
    \begin{equation}
        \hat{h}^{\mu\nu}(x) = \sum_{s=1}^2 \int\frac{ d^3 \bm k}{(2\pi)^{\frac{3}{2}}} \frac{\epsilon^{\mu\nu}(\bm k,s)}{\sqrt{2|\bm k|}} \left(\hat{a}_{\bm k,s}^\dagger e^{-i k\cdot x} + \hat{a}_{\bm k,s}e^{i k\cdot x}\right), 
        \label{hmunu}
    \end{equation}
where $\hat{a}_{\bm k}$ and $\hat{a}_{\bm k}^\dagger$ are the creation and annihilation operators and  $\epsilon^{\mu\nu}(\bm k,s)$ are the polarization tensors. As a consequence of the traceless transverse gauge, the polarization tensors are traceless and orthogonal to both the observer's four-velocity $u^\mu$ and to the corresponding momentum modes $k^\mu$. These tensors also satisfy the completeness relation
\begin{equation}
    \sum_{s=1}^2 \epsilon^{\alpha\beta}(\bm k,s) \epsilon^{\mu\nu}(\bm k,s) \!=\! \frac{1}{2}\!\left(\Pi^{\mu \alpha} \Pi^{\nu \beta}\!+\!\Pi^{\nu \alpha} \Pi^{\mu \beta}\!-\!\Pi^{\mu \nu} \Pi^{\alpha \beta}\right)\!,
\end{equation}
where $\Pi^{\alpha\beta}$ are the projectors in the space orthogonal to $u^\mu$ and the space part of $k^\mu$. The projectors are explicitly given by
\begin{equation}
    \Pi^{\alpha\beta} = \eta^{\alpha\beta} - k_\perp^\alpha k_{\perp}^\beta + u^{\alpha}u^\beta,
\end{equation}
where $k_{\perp}^\mu$ denotes the projection of $k^\mu$ in the rest space of $u^\mu$.
In particular, in a basis such that the timelike vector is chosen as $u^\mu$, the only nonzero components of $h_{\mu\nu}$ and the polarization vectors are $h_{ij}$ and $\epsilon_{ij}$, $i,j = 1,2,3$. 

The canonical commutation relations for $\hat{h}_{\mu\nu}$ and its conjugate momentum imply the usual bosonic commutation relations for the creation and annihilation operators,
\begin{equation}
    \comm{\hat{a}_{\bm k,s}}{\hat{a}^\dagger_{\bm k',s'}} = \delta_{ss'}\delta^{(3)}(\bm k - \bm k').
\end{equation}
The Fock space of the theory can then be built from a vacuum state $\ket{0}$, defined by the property
\begin{equation}
    \hat{a}_{\bm k,s}\ket{0} = 0 \:\:\:\:\: \forall\: \bm k \:\:\: \& \:\: s = 1,2.
\end{equation}
From the vacuum state, it is then possible to build the Fock space of the theory by repeated applications of the creation operators. It is usual to call states of the form $\ket{\bm k,s} \coloneqq \hat{a}^\dagger_{\bm k,s}\ket{0}$ Fock states. These states are commonly employed in the literature to describe particle states and to compute decay rates and cross sections. However, it must be noted that these states are unphysical in a number of ways. Indeed, not only are these states completely delocalized in position and not normalized ($\braket{\bm k}{\bm k'} = \delta^{(3)}(\bm k-\bm k')$), but the expected values of physically meaningful observables (such as energy) on these states is infinite\footnote{The use of these states to compute finite values of observable quantities relies on a division by their infinite norm, which yields a well defined limit~\cite{peskin,gravDetec}.}. For these reasons, instead of considering Fock states for the computation of measurable quantities, we will consider wavepackets. That is, we consider a ``smeared'' version of the Fock states that localizes them in momentum and position and yields a normalized state. A general normalized one-particle state $\ket{\psi}$ in the linearized quantum gravity theory can be written as
\begin{equation}\label{packet}
    \ket{\psi} = \sum_{s=1}^2\int \dd^3 k f_s(\bm k) a_{\bm k,s}^\dagger \ket{0},
\end{equation}
where the condition that $\ket{\psi}$ is normalized implies that $\norm{f_1}_2^2
+\norm{f_2}_2^2 = 1$, where $\norm{\:\cdot\:}_2$ denotes the $L^2$ norm. The state $\ket{\psi}$ then represents a superposition of states with different polarizations (associated with the $s$ index) with momentum profile given by the shape of the functions $f_s(\bm k)$. The normalization of the state then ensures that the expected values of valid field observables is finite, as opposed to what happens when one considers Fock states. 

As a simple example, consider the operator that corresponds to the energy of the gravitational field: the normal ordered Hamiltonian, given in terms of creation and annihilation operators by
\begin{equation}
    :\!\!\hat{H}\!\!: = \sum_{s=1}^2\int \dd^3 k \,|\bm k|\, \hat{a}_{\bm k,s}^\dagger \hat{a}_{\bm k,s}.
\end{equation}
It is easy to see that the expected value of the observable above in a Fock state $\ket{\bm k,s}$ is ill defined. However, considering a state of the form of Eq. \eqref{packet}, the expected value of the Hamiltonian takes the shape
\begin{equation}
     \langle:\!\!\hat{H}\!\!:\rangle_\psi = \sum_{s=1}^2\int \dd^3 k \,|\bm k|\, |{f}_{s}(\bm k)|^2.
\end{equation}
In this case, the energy of the state is well defined, {apart} from standard domain issues that are present even in non-relativistic quantum mechanics. With the example above it should become clear that in order to assign a finite energy to a state in quantum field theory, one must smear the Fock states in order to obtain wavepackets.

\section{The Linear Gravity-Matter Interaction}\label{sec:gMatter}

The interaction between our detector (the hydrogen atom) and the gravitational field will be based on the interaction Hamiltonian density
\begin{equation}
\mathcal{H}\approx \frac{m}{2}R_{0i0j}x^ix^j,
\label{hamiltonian density}
\end{equation}
where $m$ is the electron mass and $x^i$ denote the space Fermi normal coordinates in the atom's rest space. We motivate this interaction by discussing two different approaches having the same Hamiltonian density given by Eq.~(\ref{hamiltonian density}) as their final result.

In the first approach (considered in detail in Ref.~\cite{gravDetec}), we consider the Lagrangian interaction between the atom and the gravitational field. The interaction Lagrangian density is given by
\begin{equation}
\mathcal{L}=\frac{1}{2}h_{\mu\nu}T^{\mu\nu},
\end{equation}
where $T^{\mu\nu}$ is the energy-momentum tensor of the atom. When the typical velocities are negligible, the most relevant contribution comes from the mass-energy density $T^{00} = m$ and the Hamiltonian $H=pv-L$ is well approximated  $H=-L$. In coordinates adapted to the atom's frame and in the weak field limit it can be easily shown that $h_{00}=-R_{0i0j}x^{i}x^j$ so that we recover Eq.~(\ref{hamiltonian density}) in this case. 

We can also obtain Eq.~(\ref{hamiltonian density}) by considering the Schr\"odinger equation in a curved spacetime as a limit of the Dirac equation~\cite{jonas}. Dirac's formalism can be extended to curved spacetime, where the geometric information is encoded in the spinors. In this set up, it is expected that new terms based on the nontrivial geometry of the background spacetime survive in the  nonrelativistic Hamiltonian obtained in this limit. For an hydrogen atom following an inertial trajectory, the interaction Hamiltonian becomes~\cite{jonas} 
\begin{equation}
\mathcal{H}=\frac{m}{2}R_{0i0j}x^{i}x^j,
\end{equation}
where $R_{0i0j}$ is calculated along the hydrogen atom worldline. 

In both cases,  Eq.~(\ref{hamiltonian density}) was derived using Fermi normal coordinates. However, $h_{\mu\nu}$ is more easily expressed (and interpreted) in the traceless transverse gauge. As discussed in Ref.~\cite{gravDetec}, to lowest order in $h_{\mu\nu}$ the curvature tensor $R_{\alpha\beta\gamma\delta}$ depends on the second derivatives of $h_{\mu\nu}$ and can be expressed in the following gauge invariant expression
\begin{equation}
    R_{\alpha \beta \gamma \delta}=\frac{1}{2}\left(h_{\alpha \delta, \beta \gamma}+h_{\beta \gamma, \alpha \delta}-h_{\beta \delta, \alpha \gamma}-h_{\alpha \gamma, \beta \delta}\right).
\end{equation}
In the traceless transverse gauge the only non zero components in expression above are $R_{0i0j}$ and these can be written as
\begin{equation}\label{R0i0j}
    R_{0i0j} = -\frac{1}{2}h^{TT}_{ij,00},
\end{equation}
allowing us to rewrite the interaction in terms of our results from Section \ref{sec:linQG}.

We proceed by rewriting the interaction Hamiltonian density in terms of the eigenstates of the free Hydrogen atom. Considering the metric perturbation to be quantized, we have
\begin{equation}
\mathcal{H}=\frac{m}{2}\hat{R}_{0i0j}(t,\bm{x})x^ix^j,
\label{interaction hamiltoninan}
\end{equation}
where the curvature operator is given by Eq. \eqref{R0i0j} with the expression for the quantized perturbation given in Eq. \eqref{hmunu}. We now insert the identities $\openone =\Sigma_{n,l,m} {|nlm\rangle\langle nlm|}$ (for the hydrogen atom eigenstates) and $\openone=\int{d^3x|\bm{x}\rangle\langle \bm{x}|}$ so that we obtain 
\begin{equation}\begin{aligned}
\hat{H}_{\textrm{int}}=&\frac{m}{2}\sum_{n,l,m}\sum_{\tilde{n}\tilde{l}\tilde{m}}\int{d^3x\Bigg[\psi_{nlm}^{\ast}(t,\bm{x})x^{j}x^{j}\psi_{\tilde{n}\tilde{l}\tilde{m}}(t,\bm{x})}\\&\:\:\:\:\:\:\:\:\:\:\:\:\:\:\:\:\:\:\:\:\:\:\:\:\:\:\:\:\:\:\:\:\:\:\:\:\:\:\:\:\:\:\:\:\times \hat{R}_{0i0j}(t,\bm{x})|nlm\rangle\langle \tilde{n}\tilde{l}\tilde{m}|\Bigg].
\end{aligned}
\label{interaction picture}
\end{equation}

There are two more steps to reach our model for the interaction of the Hydrogen atom with the linear gravitational field. First we restrict ourselves to two levels $|1\rangle\equiv |nlm\rangle$ and $|2\rangle\equiv |\tilde{n}\tilde{l}\tilde{m}\rangle$ of the hydrogen atom in the interaction picture. Second, we insert a switching function $\chi(t)$ that controls the time duration of the interaction. With these we obtain
\begin{align}
\hat{H}_{\textrm{I}}(t)=&\frac{m}{2}\chi(t)\int{d^3x\Bigg[\psi_{1}^{\ast}(\bm{x})x^{j}x^{j}\psi_{2}(\bm{x})}
\label{interaction hamiltoninan}\\&\:\:\:\:\:\:\:\:\:\:\:\:\:\:\:\:\:\:\:\:\:\:\:\:\:\:\:\:\:\times \hat{R}_{0i0j}(t,\bm{x})e^{i\Omega t}|2\rangle\langle1|+\textrm{H.c.}\Bigg],\nonumber
\end{align}
where $\Omega$ is the energy gap between the levels labeled by $1$ and $2$. 

To leading order, the transition probability between the states $\ket{1}$ and $\ket{2}$ when the gravitational field is in a state $\ket{\psi}$ can be written as a function of the two-point function of the curvature operator
\begin{align}
    p_{1\rightarrow 2} = \frac{\:m^2}{4}\int \dd^4 x \dd^4 x' &\Lambda^{kl*}(x) \Lambda^{ij}(x') e^{i \Omega(t-t')}\\
    &\times\bra{\psi}\! \hat{R}_{0i0j}(x') \hat{R}_{0k0l}(x)\!\ket{\psi}\nonumber,
\end{align}
where
\begin{equation}
    \Lambda^{ij}(x) \coloneqq \chi(t)\psi_{1}^{\ast}(\bm{x})x^{i}x^{j}\psi_{2}(\bm{x})
\end{equation}
is usually called the spacetime smearing function and the curvature associated with the metric perturbation by Eq. \eqref{R0i0j} is regarded as a quantum field, given by
\begin{equation}\label{R0i0jQ}
    \hat{R}_{0i0j}(x) =\! \sum_{s=1}^2 \!\int\frac{ d^3 \bm k}{8\pi^{\frac{3}{2}}}|\bm k|^{\frac{3}{2}}\!\! \left(\hat{a}_{\bm k,s}^\dagger e^{-i k\cdot x} + \hat{a}_{\bm k,s}e^{i k\cdot x}\right)\epsilon_{ij}(\bm k,s).
\end{equation}

Assuming $\ket{\psi}$ to be a general one-particle state as in Eq. \eqref{packet}, the transition probability can be written as a sum of three terms, associated with the vacuum contribution $p_{\text{vac}}$, and the particle contributions from the rotating and counter-rotating terms:
\begin{equation}\label{prob}
    p_{1\rightarrow 2} = p_{\text{vac}}(\Omega) + p_{\text{part}}(\Omega) + p_{\text{part}}(-\Omega).
\end{equation}
These terms can be written in terms of the mode expansion of Eq. \eqref{R0i0jQ}. They read
\begin{align}
    &p_{\text{vac}}(\Omega) \!=\! \frac{m^2}{4}\!\sum_{s=1}^2 \!\int\!\! \dd^3 \bm k |\bm k|^3 |Q^{ij}(\bm k)\epsilon_{ij}(\bm k,s)|^2|\tilde{\chi}(|\bm k|\!+\!\Omega)|^2,\nonumber\\
    &\!\!p_{\text{part}}(\Omega) \!= \frac{m^2}{4}\!\left|\sum_{s=1}^2 \!\int \!\dd^3 \bm k |\bm k|^\frac{3}{2}\!f_s(\bm k)Q^{ij}(\bm k)\epsilon_{ij}(\bm k,s)\tilde{\chi}(|\bm k|\!+\!\Omega)\right|^2\!\!,
\end{align}
where $\tilde{\chi}(\omega)$ is the Fourier transform of the switching function and $Q^{ij}(\bm k)$ is the Fourier transform of the quadrupole density:
\begin{align}
    \tilde{\chi}(\omega) &= \int \dd t \chi(t) e^{-i\omega t}\\
    Q^{ij}(\bm k) &= \int \dd^3 \bm x\, \psi_1^*(\bm x) x^i x^j \psi_2(\bm x) \frac{e^{-i \bm k \cdot \bm x}}{8 \pi^{\frac{3}{2}}}.
\end{align}

The decomposition of the transition probability as in Eq. \eqref{prob} is particularly useful to identify the vacuum contribution to the excitation and make the particle contribution explicit. The $\Omega$ dependence also allows one to easily identify the rotating term (evaluated at $-\Omega$) and the counter-rotating terms (evaluated at $\Omega$). In the long-time limit, the counter-rotating terms can be shown to vanish. In particular, this implies that
\begin{equation}
    \lim_{\chi\rightarrow 1} p_{1\rightarrow 2} = p_{\text{part}}(-\Omega),
\end{equation}
where the limit above denotes the adiabatic limit, and is the most adequate way of handling interactions infinitely long, for it takes care of spurious divergences~\cite{Satz_2007,LoukoCurvedSpacetimes,erickson}.

A very important feature of most particle detector models is that for finite time interactions, the counter rotating terms do contribute. In particular, there is a non trivial excitation probability for the detector even when the field starts in the vacuum state. This excitation can be associated with the energy that has to be put into the system to switch the interaction on and off, and does not represent a physical detection of any particle content. 
Nevertheless, even in finite time interactions, the decomposition in Eq. \eqref{prob} is well defined and one can claim to have detected a particle when they measure a deviation from the standard vacuum excitation $p_{\text{vac}}(\Omega)$ that is originated by the switching. In fact, this is part of the idea that will be used in the next section to propose an angular momentum based graviton detection mechanism.

\section{Angular Momentum Based Excitations}\label{sec:main}

    We now have all the tools required to compute the transition probabilities. We will be interested in the interaction of a Hydrogen atom with the gravitational field when it starts in the $2p$ state with angular momentum in the negative $z$ direction. In principle, it is possible for the Hydrogen atom to get excited by means of this interaction. However, we will only consider transitions between states with same energy. That is, we will only consider transitions between the $n=2$ states. The eigenfunctions of the Hydrogen atom Hamiltonian for $n=2$ read
    \begin{align}
        \psi_{2,0,0}(r) &= \frac{1}{4\sqrt{2 \pi} a_0^\frac{3}{2}}(2-r/a_0)e^{-\frac{r}{2a_0}},\\
        \psi_{2,1,0}(r,\theta) &= \frac{1}{4\sqrt{2 \pi} a_0^\frac{3}{2}} \frac{r}{a_0}e^{-\frac{r}{2a_0}} \cos\theta,\\
        \psi_{2,1,\pm 1}(r,\theta,\phi) &= \frac{1}{4\sqrt{2 \pi} a_0^\frac{3}{2}} \frac{r}{a_0}e^{-\frac{r}{2a_0}}e^{\pm i \phi} \sin\theta,
    \end{align}
    \textcolor{black}{where $a_0$ is the Bohr radius.} We will be interested in transition probabilities when the detector starts in the state $\psi_{2,1,-1}$ and changes to any of the states above. These transitions are then associated with exchanges of angular momentum only and do not require the detector to exchange energy with the gravitational field. Instead, the energy exchange is mediated by the switching of the interaction. Notice that due to the fact that all states considered posses a symmetry around the $z$-axis, angular momentum in the $z$-direction is not affected by the switching of the detector, and any exchange of angular momentum must come from the gravitational field.
    
    Given that the states we are interested in all have the same energy, it can be shown that the transition probabilities vanish in the long time limit due to the vanishing effective energy gap in Eq. \eqref{interaction hamiltoninan}.  Therefore,  in order to obtain nontrivial probabilities, we will consider finite time interactions prescribed by the Gaussian switching function
    \begin{equation}
        \chi(t) = \frac{1}{\sqrt{2\pi}}e^{-\frac{t^2}{2T^2}}.
    \end{equation}
    With the above choice, we have the approximate time duration of the interaction given by $T$. The switching function's Fourier transform then reads
    \begin{equation}
        \tilde{\chi}(k) = T e^{-\frac{k^2T^2}{2}}.
    \end{equation}
    
    In order to compute the transition probabilities explicitly, we must parametrize the polarization tensors $\epsilon_{\mu\nu}$ in Eq. \eqref{hmunu}. Choosing $k^\mu = (1,0,0,1)$ in a given inertial frame yields the following expression for the polarization tensors in the associated basis
    \begin{align} 
        \epsilon\,(\bm k,1) = \frac{1}{\sqrt{2}}\!\!\begin{pmatrix}
            0 & 0 & 0 & 0\\
            0 & 1 & 0 & 0\\
            0 & 0 & -1 & 0\\
            0 & 0 & 0 & 0
        \end{pmatrix},\:
        \epsilon\,(\bm k,2) = \frac{1}{\sqrt{2}}\!\!\begin{pmatrix}
            0 & 0 & 0 & 0\\
            0 & 0 & 1 & 0\\
            0 & 1 & 0 & 0\\
            0 & 0 & 0 & 0
        \end{pmatrix}.
    \end{align}
    More generally, given two spacelike vectors $e_1(\bm k)$ and $e_2(\bm k)$ such that $u_\mu e^\mu_i = k_\mu e^\mu_i = 0$ for $i = 1,2$, we can choose the polarization vectors as
    \begin{align}
        \epsilon\,(\bm  k,1) = \frac{1}{\sqrt{2}}\left(e_1(\bm k)\otimes e_1(\bm k) - e_2(\bm k) \otimes e_2(\bm k)\right),\label{e1}\\
        \epsilon\,(\bm k,2) = \frac{1}{\sqrt{2}}\left(e_1(\bm k)\otimes e_2(\bm k) + e_2(\bm k) \otimes e_1(\bm k)\right).\label{e2}
    \end{align}
    It is possible to parametrize the integration space in \eqref{hmunu} according to spherical coordinates $(|\bm k|,\alpha,\psi)$, so that we can write $\bm k$ and define the vectors $\bm e_1$ and $\bm e_2$ according to
        \begin{align}
        \bm k &= |\bm k| (\sin\alpha \cos\psi \: \bm e_x + \sin \alpha \sin \psi \: \bm e_y + \cos \alpha \bm e_z ),\nonumber\\
        \bm e_1(\alpha,\psi) &=  \cos\alpha \cos\psi \: \bm e_x + \cos \alpha \sin \psi \: \bm e_y - \sin \alpha \bm e_z ,\nonumber\\
        \bm e_2(\alpha,\psi) &= -\sin\psi \: \bm e_x + \cos \psi \: \bm e_y.
        \end{align}
    Plugging the expressions above into Eqs. \eqref{e1} and \eqref{e2} gives the following expressions for the \textcolor{black}{spatial part of the polarization tensors},
    \begin{widetext}
    \begin{equation}
        \epsilon_1(\alpha,\psi) = \frac{1}{\sqrt{2}}\left(
         \begin{array}{ccc}
             \cos ^2\alpha  \cos ^2\psi -\sin ^2\psi  & \cos ^2\alpha  \sin \psi \cos \psi+\sin \psi  \cos \psi  & -\sin \alpha  \cos
               \alpha  \cos \psi  \\
             \cos ^2\alpha  \sin\psi  \cos \psi+\sin\psi  \cos \psi  & \cos ^2\alpha  \sin ^2\psi -\cos ^2\psi  & -\sin \alpha\cos
               \alpha  \sin \psi  \\
            - \sin \alpha\cos \alpha \cos \psi  & -\sin \alpha\cos\alpha \sin\psi & \sin ^2\alpha  \\
        \end{array}
        \right),
    \end{equation}
    \begin{equation}
        \epsilon_2(\alpha,\psi) = \frac{1}{\sqrt{2}}\left(
            \begin{array}{ccc}
             -2 \cos \alpha  \sin \psi  \cos \psi  & \cos \alpha \cos ^2\psi -\cos \alpha  \sin ^2\psi & \sin \alpha  \sin \psi \\
             \cos\alpha \cos^2\psi-\cos \alpha \sin^2\psi & 2 \cos \alpha \sin\psi \cos\psi & -\sin \alpha \cos \psi \\
             \sin \alpha ) \sin \psi & -\sin \alpha \cos \psi  & 0 \\
            \end{array}
            \right).
    \end{equation}
    \end{widetext}
    
    The next step to compute the transition probabilities is to input the information regarding the initial state of the atom in Eq. \eqref{interaction hamiltoninan}. That is, we must compute the interaction's quadrupole moment densities and their Fourier transforms.
    The Fourier transform of any quadrupole can be rewritten in simpler terms as the second derivative of a scalar Fourier transform:
    \begin{equation}\label{easy}
        Q_{ij}(\bm k) = - \pdv{}{k^i}\pdv{}{k^j}q(\bm k),
    \end{equation}
    where
    \begin{equation}
        q(\bm k) = \frac{1}{8\pi^\frac{3}{2}}\int \dd^3 \bm x\, \psi_1^*(\bm x)\psi_2(\bm x) e^{-i \bm k \cdot \bm x}.
    \end{equation}
    The function $q(\bm k)$ for the processes we are considering can be computed analytically by performing an adequate change of variables which aligns $\bm k$ to the $z$-axis. Denoting by $q_0$, $q_1$ and $q_2$ the functions associated with the transitions $\psi_{2,1,-1}\rightarrow \psi_{2,0,0}$, $\psi_{2,1,-1}\rightarrow \psi_{2,1,0}$ and $\psi_{2,1,-1}\rightarrow \psi_{2,1,1}$, respectively, we have, in spherical coordinates $\bm k = (|\bm k|,\alpha,\psi)$,
    \begin{align}
        q_{0}(\bm k) &=  \frac{3 i |\bm k|(1-|\bm k|^2) e^{i \psi}\sin\alpha}{16\pi^\frac{3}{2}(1+|\bm k|^2)^4},\\
        q_{1}(\bm k) &=  \frac{3 |\bm k|^2 e^{i \psi}\sin\alpha\cos\alpha}{16\pi^\frac{3}{2}(1+|\bm k|^2)^4},\\
        q_{2}(\bm k) &=  \frac{3 |\bm k|^2 e^{2i \psi}\sin^2\alpha}{16\pi^\frac{3}{2}(1+|\bm k|^2)^4}.
    \end{align}
    Notice how the difference between the angular momentum of the different states is associated with the azimuthal angle dependence of the Fourier transforms above: $e^{in\psi}$ is associated with a difference in angular momentum of $\hbar n$ between the states considered. 
    
    With these preliminary results, it is possible to show that the transition to the $2s$ state, $\psi_{2,1,-1}\rightarrow \psi_{2,0,0}$, is not allowed. That is, $Q^{ij}_0(\bm k)\epsilon_{ij}(\bm k,s) = 0$ for $s=1,2$ and the transition probability vanishes. However, the transitions to the other $2p$ states are non trivial. 
    
    The vacuum contributions for these processes is the same and can be found analytically,
    \begin{align}
        p_{\text{vac}}({\color{black}\Omega = 0}) =& \frac{T^2{\color{black} m^2a_0^2}}{44800} \Big[e^{T^2}\!\! \left(T^4+14 T^2+42\right) T^{10} \text{Ei}\left(-T^2\right)\nonumber\\
        &\!\!\!\!\!\!\!\!\!\!\!\!\!\!\!\!\!\!\!\!\!\!\!\!\!\!+48+T^{12}+13 T^{10}+30 T^8-20 T^6+24 T^4-36
   T^2\Big].
    \end{align}
In order to compute the particle state excitation to the probability, we must choose the momentum space profile functions that define the  wavepacket of Eq. \eqref{packet}. However, it must be noted that not all particle states will contribute nontrivially: They must contain enough angular momentum in the $z$ direction to promote the corresponding excitations. We then consider states with the following choices of momentum distributions
    \begin{align}
        f_1(\bm k) &= \frac{\alpha_1}{(\pi\sigma^2)^{\frac{3}{4}}}e^{-\frac{|\bm k-\bm k_0|^2}{2\sigma^2}}e^{-i {\color{black} l } \psi},\\  f_2(\bm k ) &= \frac{\alpha_2}{(\pi\sigma^2)^{\frac{3}{4}}}e^{-\frac{|\bm k-\bm k_0|^2}{2\sigma^2}}e^{-i {\color{black} l} \psi},
    \end{align}
    where $\bm k_0$ is the center of the momentum distribution, $\sigma$ is its width and {\color{black}$l$} corresponds to the angular momentum of the state. The normalization condition for the state $\ket{\psi}$ implies that the coefficients $\alpha_1$ and $\alpha_2$ satisfy $|\alpha_1|^2 + |\alpha_2|^2 = 1$. With the above choices of momentum profile functions, it can be shown that the average energy in the state $\ket{\psi}$ with $\bm k_0=0$ is given by
    \begin{equation}
        \langle:\!\!\hat{H}\!\!:\rangle_{\psi} = \frac{2\sigma}{\sqrt{\pi}},
    \end{equation}
    so that in this case $\sigma$ is not only the width of the distribution around $\bm k_0 = \bm 0$, it is also proportional to the energy of the wavepacket.
    
\begin{figure}[h!]
        \centering
        \includegraphics[scale=0.67]{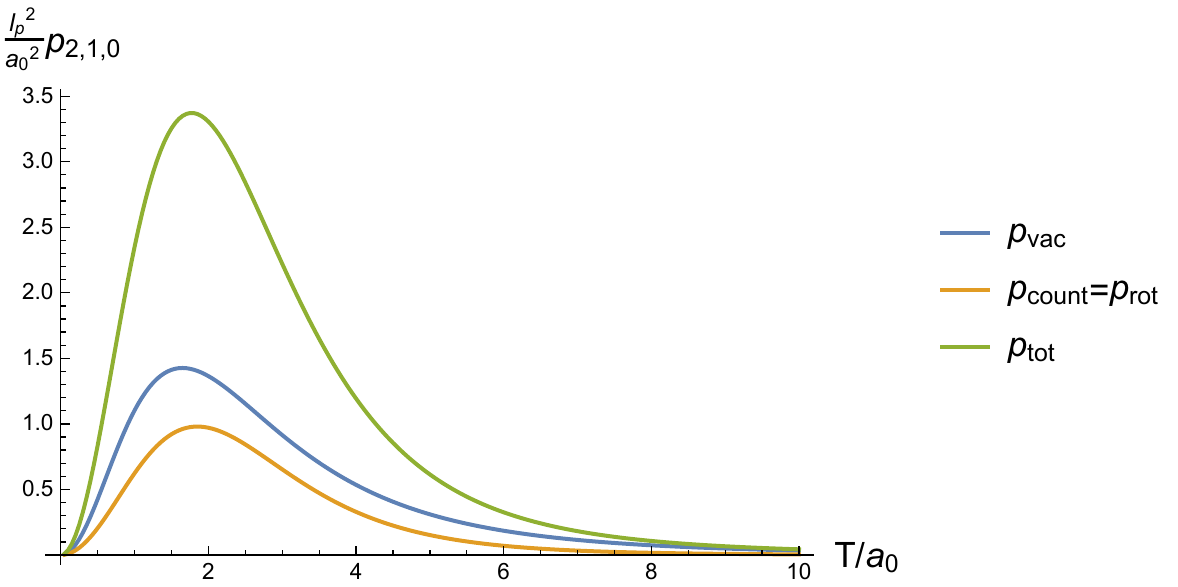}
        \caption{Vacuum, particle and total transition probability for the process $\psi_{2,1,-1}\rightarrow \psi_{2,1,0}$ for an atom interacting with the linearized quantum gravitational field as a function of the average interaction time $T$. We have chosen the Gaussian state from Eq. \eqref{packet} with $\alpha_1 = 0$, $\alpha_2=1$, $\sigma = 0.4 \:a_0^{-1}$ and $\bm k_0 = 0$.}
        \label{fig:0}
    \end{figure}

\begin{figure}[h!]
        \centering
        \includegraphics[scale=0.67]{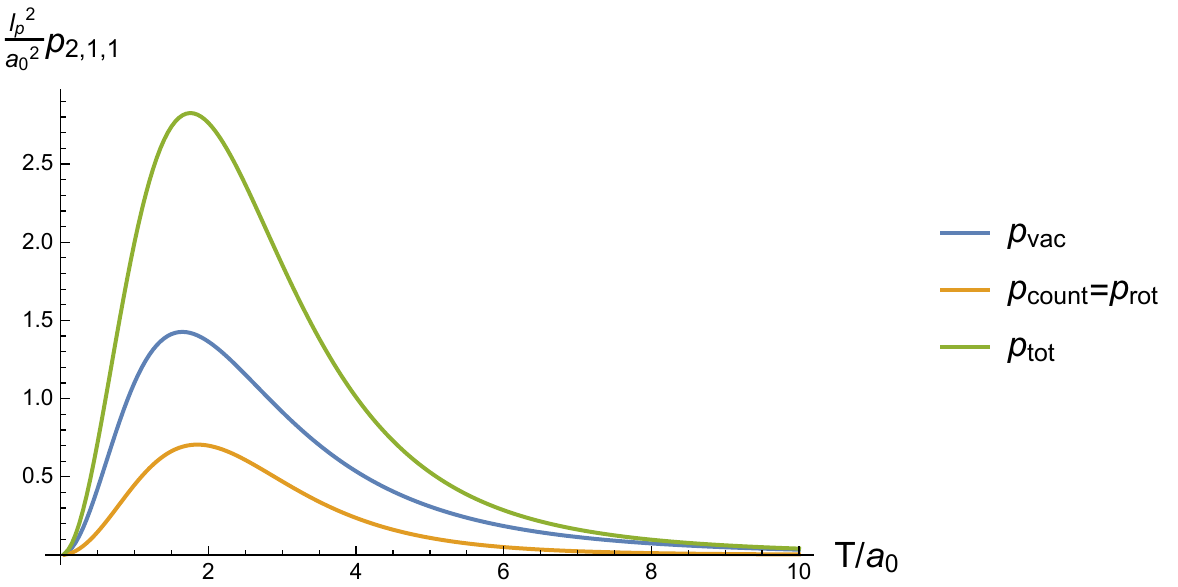}
        \caption{Vacuum, particle and total transition probability for the process $\psi_{2,1,-1}\rightarrow \psi_{2,1,1}$ for an atom interacting with the linearized quantum gravitational field as a function of the average interaction time $T$. We have chosen the Gaussian state from Eq. \eqref{packet} with $\alpha_1 = 1$, $\alpha_2=0$, $\sigma = 0.4 \:a_0^{-1}$ and $\bm k_0 = 0$.}
        \label{fig:1}
    \end{figure}

    In order to obtain non-vanishing transition probabilities associated with the states above, we choose ${\color{black} l}=1$ in the case of the $\psi_{2,1,-1}\rightarrow \psi_{2,1,0}$ transition and ${\color{black} l }=2$ in the case of $\psi_{2,1,-1}\rightarrow \psi_{2,1,1}$. With these and the choices of $\sigma = 0.4$ and $\bm k_0 = \bm 0$, we obtain the plots from Fig. \ref{fig:0} and Fig. \ref{fig:1}.
    
    Our choices for the gravitational field states  their average energy to $E \approx 1.65\text{eV}$, which is less than the energy gap from the $n=2$ to the $n=3$ orbitals, $E_3-E_2 \approx 1.89\text{eV}$. In both cases we can see that the effect of the particle state on the excitation probability is of the same order of magnitude as the vacuum contribution, so that its effect is, in principle, measurable, even though the energy of the gravity states is less than the energy gap of the particle detector. With this, we have an angular momentum based particle detector, that does not rely on an energy exchange with the gravitational field, but rather, relies solely in an angular momentum exchange. In the next section we will discuss how to implement the finite time interaction and how to interpret our results.
    
\section{Implementing the finite time interaction}\label{implementing}

In this section we discuss the implementability of the method discussed in Section \ref{sec:main}. Notice that Figs.~\ref{fig:0} and~\ref{fig:1} show that the total probabilities for the transitions $\psi_{2,1,-1}\rightarrow \psi_{2,1,0}$ and $\psi_{2,1,-1}\rightarrow \psi_{2,1,1}$ vanish in the $T\to\infty$ limit.  This is expected since  in the infinite time interaction  the transient effects turn out to be  irrelevant and the considered transitions becomes forbidden on energy conservation grounds. Therefore,  the finite time interaction has a crucial role in our angular momentum based detector.

To implement this finite time interaction we first consider an ensemble of hydrogen atoms in the ground state $\psi_{1,0,0}$. These atoms can then be excited through the interaction with polarized photons with energy $E=E_{n=2}-E_{n=1}$ and spin $-\hbar$. This process will give an ensemble of hydrogen atoms in the $\psi_{2,1,-1}$ state, effectively turning on the interaction with the gravitational field considered in the previous section, where the detector starts in the $\psi_{2,1,-1}$ state. The excited electron will stay in this state for a short time so that the mean life-time of the hydrogen $2p$ state gives us the time duration of the interaction.

In this description, the interaction with the electromagnetic field in turned on indefinitely, which sweeps any possible vacuum contribution that might come from this interaction. When only the electromagnetic interaction is considered, the final state will be composed of atoms in the ground states and polarized (spin down) photons. However, as we have shown in the previous section, the interaction with the gravitational field may produce hydrogen atoms in  $\psi_{2,1,0}$ and $\psi_{2,1,1}$ states so that, in the end, photons with energy $E=E_{n=2}-E_{n=1}$ can be re-emitted with positive angular momentum. In this way, the presence of positive angular momentum photons as final states can then be used to indicate that gravitons interacted with the detector in order to provide it with angular momentum. It is important to notice that the mere interaction with the vacuum of the gravitational field will provoke excitations, as seen in Figs. \ref{fig:0} and \ref{fig:1}. These excitations correspond to the deposit of angular momentum in the gravitational field by means of emissions of gravitons. The energy for the creation of these particles then comes from the switching of the interaction. However, in the plots of Figs. \ref{fig:0} and \ref{fig:1}, we see that there are graviton states whose effect in the transition probability is of the same order as the vacuum contribution. So, in principle, it is possible to compare the amount of photons with positive angular momentum with the expected amount from the vacuum contribution. Any meaningful difference will then indicate the detection of graviton particles.

This effect then allows one to detect gravitons with energies smaller than $E_{n=3}-E_{n=2}$. Moreover, this procedure could be adapted to atoms with electrons in the outermost shell in order to measure gravitons with even smaller energies.

Finally, it is important to remark that although our framework provides a new approach for the detection of gravitons, the order of magnitude of the leading order transition probabilities is of the order of $\ell_p^2/a_0^2\approx 10^{-48}$, where $\ell_p$ is the Planck length. This implies that with current experimental technology, this effect is still cannot be measured. However, it provides an approach to graviton detection based on apparatus sensitivity only, regardless of the energy of the graviton particles involved in the process.

\section{Conclusions}\label{sec:conclusion}

Particle detectors in  general curved spacetimes are based on a non-relativistic system which couples to a quantum field propagating in the classical background.  The detector is said to click when there is a transition between two energy levels, which is characterized by the energy gap $\Omega$. We then we say that a particle has been detected.  The energy of the  detected particle must be at least equal to $\Omega$ so that lower energy particles turn out to be  invisible to the detector.

In this paper we  presented a mechanism in which low energy particles can be detected.  We illustrate this mechanism by probing the quantized gravitational field in the linear approximation.  By considering a one-particle smeared state  to represent the graviton, we showed  that the difference between the transition probability for this one particle state and pure transient effects given by the vacuum contribution indicates the graviton detection.

\acknowledgments

The authors would like to thank Ricardo A. Mosna,  George E. A. Matsas and Leonardo P. de Gioia for insightful discussions. T. R. P. also thanks Drs. David Kubiz\v{n}\'ak and  Eduardo Mart\'in-Mart\'inez’s funding through their NSERC Discovery grants. Research at Perimeter Institute is supported in part by the Government of Canada through the Department of Innovation, Science and Industry Canada and by the Province of Ontario through the Ministry of Colleges and Universities.

\bibliography{references}

\begin{thebibliography}{39}%
\makeatletter
\providecommand \@ifxundefined [1]{%
 \@ifx{#1\undefined}
}%
\providecommand \@ifnum [1]{%
 \ifnum #1\expandafter \@firstoftwo
 \else \expandafter \@secondoftwo
 \fi
}%
\providecommand \@ifx [1]{%
 \ifx #1\expandafter \@firstoftwo
 \else \expandafter \@secondoftwo
 \fi
}%
\providecommand \natexlab [1]{#1}%
\providecommand \enquote  [1]{``#1''}%
\providecommand \bibnamefont  [1]{#1}%
\providecommand \bibfnamefont [1]{#1}%
\providecommand \citenamefont [1]{#1}%
\providecommand \href@noop [0]{\@secondoftwo}%
\providecommand \href [0]{\begingroup \@sanitize@url \@href}%
\providecommand \@href[1]{\@@startlink{#1}\@@href}%
\providecommand \@@href[1]{\endgroup#1\@@endlink}%
\providecommand \@sanitize@url [0]{\catcode `\\12\catcode `\$12\catcode
  `\&12\catcode `\#12\catcode `\^12\catcode `\_12\catcode `\%12\relax}%
\providecommand \@@startlink[1]{}%
\providecommand \@@endlink[0]{}%
\providecommand \url  [0]{\begingroup\@sanitize@url \@url }%
\providecommand \@url [1]{\endgroup\@href {#1}{\urlprefix }}%
\providecommand \urlprefix  [0]{URL }%
\providecommand \Eprint [0]{\href }%
\providecommand \doibase [0]{http://dx.doi.org/}%
\providecommand \selectlanguage [0]{\@gobble}%
\providecommand \bibinfo  [0]{\@secondoftwo}%
\providecommand \bibfield  [0]{\@secondoftwo}%
\providecommand \translation [1]{[#1]}%
\providecommand \BibitemOpen [0]{}%
\providecommand \bibitemStop [0]{}%
\providecommand \bibitemNoStop [0]{.\EOS\space}%
\providecommand \EOS [0]{\spacefactor3000\relax}%
\providecommand \BibitemShut  [1]{\csname bibitem#1\endcsname}%
\let\auto@bib@innerbib\@empty
\bibitem [{\citenamefont {Weinberg}(1965)}]{weinberg}%
  \BibitemOpen
  \bibfield  {author} {\bibinfo {author} {\bibfnamefont {S.}~\bibnamefont
  {Weinberg}},\ }\href {\doibase 10.1103/PhysRev.138.B988} {\bibfield
  {journal} {\bibinfo  {journal} {Phys. Rev.}\ }\textbf {\bibinfo {volume}
  {138}},\ \bibinfo {pages} {B988} (\bibinfo {year} {1965})}\BibitemShut
  {NoStop}%
\bibitem [{\citenamefont {Weinberg}(1972)}]{weinbergBook}%
  \BibitemOpen
  \bibfield  {author} {\bibinfo {author} {\bibfnamefont {S.}~\bibnamefont
  {Weinberg}},\ }\href@noop {} {\emph {\bibinfo {title} {Gravitation and
  cosmology}}}\ (\bibinfo  {publisher} {John Wiley and Sons},\ \bibinfo
  {address} {New York},\ \bibinfo {year} {1972})\BibitemShut {NoStop}%
\bibitem [{\citenamefont {Dyson}(2004)}]{dysonReview}%
  \BibitemOpen
  \bibfield  {author} {\bibinfo {author} {\bibfnamefont {F.~J.}\ \bibnamefont
  {Dyson}},\ }\href@noop {} {\bibfield  {journal} {\bibinfo  {journal} {New
  York Review of Books}\ }\textbf {\bibinfo {volume} {51}} (\bibinfo {year}
  {2004})}\BibitemShut {NoStop}%
\bibitem [{\citenamefont {Boughn}\ and\ \citenamefont
  {Rothman}(2006)}]{gravDetec}%
  \BibitemOpen
  \bibfield  {author} {\bibinfo {author} {\bibfnamefont {S.}~\bibnamefont
  {Boughn}}\ and\ \bibinfo {author} {\bibfnamefont {T.}~\bibnamefont
  {Rothman}},\ }\href {\doibase 10.1088/0264-9381/23/20/006} {\bibfield
  {journal} {\bibinfo  {journal} {Class. Quant. Grav.}\ }\textbf {\bibinfo
  {volume} {23}},\ \bibinfo {pages} {5839} (\bibinfo {year} {2006})},\ \Eprint
  {http://arxiv.org/abs/gr-qc/0605052} {arXiv:gr-qc/0605052} \BibitemShut
  {NoStop}%
\bibitem [{\citenamefont {Rothman}\ and\ \citenamefont
  {Boughn}(2006)}]{Rothman}%
  \BibitemOpen
  \bibfield  {author} {\bibinfo {author} {\bibfnamefont {T.}~\bibnamefont
  {Rothman}}\ and\ \bibinfo {author} {\bibfnamefont {S.}~\bibnamefont
  {Boughn}},\ }\href {\doibase 10.1007/s10701-006-9081-9} {\bibfield  {journal}
  {\bibinfo  {journal} {Found. Phys.}\ }\textbf {\bibinfo {volume} {36}},\
  \bibinfo {pages} {1801} (\bibinfo {year} {2006})}\BibitemShut {NoStop}%
\bibitem [{\citenamefont {Birrell}\ and\ \citenamefont
  {Davies}(1982)}]{birrell_davies}%
  \BibitemOpen
  \bibfield  {author} {\bibinfo {author} {\bibfnamefont {N.~D.}\ \bibnamefont
  {Birrell}}\ and\ \bibinfo {author} {\bibfnamefont {P.~C.~W.}\ \bibnamefont
  {Davies}},\ }\href {\doibase 10.1017/CBO9780511622632} {\emph {\bibinfo
  {title} {Quantum Fields in Curved Space}}},\ Cambridge Monographs on
  Mathematical Physics\ (\bibinfo  {publisher} {Cambridge University Press},\
  \bibinfo {year} {1982})\BibitemShut {NoStop}%
\bibitem [{\citenamefont {Fulling}(1973)}]{fullingUnruhEffect}%
  \BibitemOpen
  \bibfield  {author} {\bibinfo {author} {\bibfnamefont {S.~A.}\ \bibnamefont
  {Fulling}},\ }\href {\doibase 10.1103/PhysRevD.7.2850} {\bibfield  {journal}
  {\bibinfo  {journal} {Phys. Rev. D}\ }\textbf {\bibinfo {volume} {7}},\
  \bibinfo {pages} {2850} (\bibinfo {year} {1973})}\BibitemShut {NoStop}%
\bibitem [{\citenamefont {Unruh}(1976)}]{Unruh1976}%
  \BibitemOpen
  \bibfield  {author} {\bibinfo {author} {\bibfnamefont {W.~G.}\ \bibnamefont
  {Unruh}},\ }\href {\doibase 10.1103/PhysRevD.14.870} {\bibfield  {journal}
  {\bibinfo  {journal} {Phys. Rev. D}\ }\textbf {\bibinfo {volume} {14}},\
  \bibinfo {pages} {870} (\bibinfo {year} {1976})}\BibitemShut {NoStop}%
\bibitem [{\citenamefont {DeWitt}(1980)}]{DeWitt}%
  \BibitemOpen
  \bibfield  {author} {\bibinfo {author} {\bibfnamefont {B.}~\bibnamefont
  {DeWitt}},\ }\href@noop {} {\emph {\bibinfo {title} {General Relativity; an
  Einstein Centenary Survey}}}\ (\bibinfo  {publisher} {Cambridge University
  Press},\ \bibinfo {address} {Cambridge, UK},\ \bibinfo {year}
  {1980})\BibitemShut {NoStop}%
\bibitem [{\citenamefont {Mart\'{\i}n-Mart\'{\i}nez}\ \emph
  {et~al.}(2013)\citenamefont {Mart\'{\i}n-Mart\'{\i}nez}, \citenamefont
  {Montero},\ and\ \citenamefont {del Rey}}]{eduardoOld}%
  \BibitemOpen
  \bibfield  {author} {\bibinfo {author} {\bibfnamefont {E.}~\bibnamefont
  {Mart\'{\i}n-Mart\'{\i}nez}}, \bibinfo {author} {\bibfnamefont
  {M.}~\bibnamefont {Montero}}, \ and\ \bibinfo {author} {\bibfnamefont
  {M.}~\bibnamefont {del Rey}},\ }\href {\doibase 10.1103/PhysRevD.87.064038}
  {\bibfield  {journal} {\bibinfo  {journal} {Phys. Rev. D}\ }\textbf {\bibinfo
  {volume} {87}},\ \bibinfo {pages} {064038} (\bibinfo {year}
  {2013})}\BibitemShut {NoStop}%
\bibitem [{\citenamefont {Mart\'{i}n-Mart\'{i}nez}\ and\ \citenamefont
  {Rodriguez-Lopez}(2018)}]{eduardo}%
  \BibitemOpen
  \bibfield  {author} {\bibinfo {author} {\bibfnamefont {E.}~\bibnamefont
  {Mart\'{i}n-Mart\'{i}nez}}\ and\ \bibinfo {author} {\bibfnamefont
  {P.}~\bibnamefont {Rodriguez-Lopez}},\ }\href {\doibase
  10.1103/PhysRevD.97.105026} {\bibfield  {journal} {\bibinfo  {journal} {Phys.
  Rev. D}\ }\textbf {\bibinfo {volume} {97}},\ \bibinfo {pages} {105026}
  (\bibinfo {year} {2018})}\BibitemShut {NoStop}%
\bibitem [{\citenamefont {Funai}\ \emph {et~al.}(2019)\citenamefont {Funai},
  \citenamefont {Louko},\ and\ \citenamefont
  {Mart\'{\i}n-Mart\'{\i}nez}}]{Nicho1}%
  \BibitemOpen
  \bibfield  {author} {\bibinfo {author} {\bibfnamefont {N.}~\bibnamefont
  {Funai}}, \bibinfo {author} {\bibfnamefont {J.}~\bibnamefont {Louko}}, \ and\
  \bibinfo {author} {\bibfnamefont {E.}~\bibnamefont
  {Mart\'{\i}n-Mart\'{\i}nez}},\ }\href {\doibase 10.1103/PhysRevD.99.065014}
  {\bibfield  {journal} {\bibinfo  {journal} {Phys. Rev. D}\ }\textbf {\bibinfo
  {volume} {99}},\ \bibinfo {pages} {065014} (\bibinfo {year}
  {2019})}\BibitemShut {NoStop}%
\bibitem [{\citenamefont {Lopp}\ and\ \citenamefont
  {Mart\'{i}n-Mart\'{i}nez}(2021)}]{richard}%
  \BibitemOpen
  \bibfield  {author} {\bibinfo {author} {\bibfnamefont {R.}~\bibnamefont
  {Lopp}}\ and\ \bibinfo {author} {\bibfnamefont {E.}~\bibnamefont
  {Mart\'{i}n-Mart\'{i}nez}},\ }\href {\doibase 10.1103/PhysRevA.103.013703}
  {\bibfield  {journal} {\bibinfo  {journal} {Phys. Rev. A}\ }\textbf {\bibinfo
  {volume} {103}},\ \bibinfo {pages} {013703} (\bibinfo {year}
  {2021})}\BibitemShut {NoStop}%
\bibitem [{\citenamefont {Torres}\ \emph {et~al.}(2020)\citenamefont {Torres},
  \citenamefont {Rick~Perche}, \citenamefont {Landulfo},\ and\ \citenamefont
  {Matsas}}]{neutrinos}%
  \BibitemOpen
  \bibfield  {author} {\bibinfo {author} {\bibfnamefont {B.~d. S.~L.}\
  \bibnamefont {Torres}}, \bibinfo {author} {\bibfnamefont {T.}~\bibnamefont
  {Rick~Perche}}, \bibinfo {author} {\bibfnamefont {A.~G.~S.}\ \bibnamefont
  {Landulfo}}, \ and\ \bibinfo {author} {\bibfnamefont {G.~E.~A.}\ \bibnamefont
  {Matsas}},\ }\href {\doibase 10.1103/PhysRevD.102.093003} {\bibfield
  {journal} {\bibinfo  {journal} {Phys. Rev. D}\ }\textbf {\bibinfo {volume}
  {102}},\ \bibinfo {pages} {093003} (\bibinfo {year} {2020})}\BibitemShut
  {NoStop}%
\bibitem [{\citenamefont {Perche}\ and\ \citenamefont
  {Martín-Martínez}(2021)}]{antiparticles}%
  \BibitemOpen
  \bibfield  {author} {\bibinfo {author} {\bibfnamefont {T.~R.}\ \bibnamefont
  {Perche}}\ and\ \bibinfo {author} {\bibfnamefont {E.}~\bibnamefont
  {Martín-Martínez}},\ }\href@noop {} {\enquote {\bibinfo {title}
  {Anti-particle detector models in qft},}\ } (\bibinfo {year} {2021}),\
  \Eprint {http://arxiv.org/abs/2106.03874} {arXiv:2106.03874 [quant-ph]}
  \BibitemShut {NoStop}%
\bibitem [{\citenamefont {Perche}()}]{mineInPreparation}%
  \BibitemOpen
  \bibfield  {author} {\bibinfo {author} {\bibfnamefont {T.~R.}\ \bibnamefont
  {Perche}},\ }\href@noop {} {\enquote {\bibinfo {title} {General features of
  thermalization of particle detectors and the unruh effect},}\ }\bibinfo
  {note} {(In Preparation)}\BibitemShut {NoStop}%
\bibitem [{\citenamefont {Gibbons}\ and\ \citenamefont
  {Hawking}(1977)}]{HawkingGibons}%
  \BibitemOpen
  \bibfield  {author} {\bibinfo {author} {\bibfnamefont {G.~W.}\ \bibnamefont
  {Gibbons}}\ and\ \bibinfo {author} {\bibfnamefont {S.~W.}\ \bibnamefont
  {Hawking}},\ }\href {\doibase 10.1103/PhysRevD.15.2738} {\bibfield  {journal}
  {\bibinfo  {journal} {Phys. Rev. D}\ }\textbf {\bibinfo {volume} {15}},\
  \bibinfo {pages} {2738} (\bibinfo {year} {1977})}\BibitemShut {NoStop}%
\bibitem [{\citenamefont {Unruh}\ and\ \citenamefont
  {Wald}(1984)}]{Unruh-Wald}%
  \BibitemOpen
  \bibfield  {author} {\bibinfo {author} {\bibfnamefont {W.~G.}\ \bibnamefont
  {Unruh}}\ and\ \bibinfo {author} {\bibfnamefont {R.~M.}\ \bibnamefont
  {Wald}},\ }\href {\doibase 10.1103/PhysRevD.29.1047} {\bibfield  {journal}
  {\bibinfo  {journal} {Phys. Rev. D}\ }\textbf {\bibinfo {volume} {29}},\
  \bibinfo {pages} {1047} (\bibinfo {year} {1984})}\BibitemShut {NoStop}%
\bibitem [{\citenamefont {Takagi}(1986)}]{Takagi}%
  \BibitemOpen
  \bibfield  {author} {\bibinfo {author} {\bibfnamefont {S.}~\bibnamefont
  {Takagi}},\ }\href {\doibase 10.1143/PTP.88.1} {\bibfield  {journal}
  {\bibinfo  {journal} {Prog. Theor. Phys. Supp.}\ }\textbf {\bibinfo {volume}
  {88}},\ \bibinfo {pages} {1} (\bibinfo {year} {1986})}\BibitemShut {NoStop}%
\bibitem [{\citenamefont {Hodgkinson}\ and\ \citenamefont
  {Louko}(2012)}]{bhDetectorsBTZ}%
  \BibitemOpen
  \bibfield  {author} {\bibinfo {author} {\bibfnamefont {L.}~\bibnamefont
  {Hodgkinson}}\ and\ \bibinfo {author} {\bibfnamefont {J.}~\bibnamefont
  {Louko}},\ }\href {\doibase 10.1103/PhysRevD.86.064031} {\bibfield  {journal}
  {\bibinfo  {journal} {Phys. Rev. D}\ }\textbf {\bibinfo {volume} {86}},\
  \bibinfo {pages} {064031} (\bibinfo {year} {2012})}\BibitemShut {NoStop}%
\bibitem [{\citenamefont {Hodgkinson}\ \emph {et~al.}(2014)\citenamefont
  {Hodgkinson}, \citenamefont {Louko},\ and\ \citenamefont
  {Ottewill}}]{bhDetectors}%
  \BibitemOpen
  \bibfield  {author} {\bibinfo {author} {\bibfnamefont {L.}~\bibnamefont
  {Hodgkinson}}, \bibinfo {author} {\bibfnamefont {J.}~\bibnamefont {Louko}}, \
  and\ \bibinfo {author} {\bibfnamefont {A.~C.}\ \bibnamefont {Ottewill}},\
  }\href {\doibase 10.1103/PhysRevD.89.104002} {\bibfield  {journal} {\bibinfo
  {journal} {Phys. Rev. D}\ }\textbf {\bibinfo {volume} {89}},\ \bibinfo
  {pages} {104002} (\bibinfo {year} {2014})}\BibitemShut {NoStop}%
\bibitem [{\citenamefont {Ng}\ \emph {et~al.}(2014)\citenamefont {Ng},
  \citenamefont {Hodgkinson}, \citenamefont {Louko}, \citenamefont {Mann},\
  and\ \citenamefont {Mart\'{\i}n-Mart\'{\i}nez}}]{bhDetectorsAdS}%
  \BibitemOpen
  \bibfield  {author} {\bibinfo {author} {\bibfnamefont {K.~K.}\ \bibnamefont
  {Ng}}, \bibinfo {author} {\bibfnamefont {L.}~\bibnamefont {Hodgkinson}},
  \bibinfo {author} {\bibfnamefont {J.}~\bibnamefont {Louko}}, \bibinfo
  {author} {\bibfnamefont {R.~B.}\ \bibnamefont {Mann}}, \ and\ \bibinfo
  {author} {\bibfnamefont {E.}~\bibnamefont {Mart\'{\i}n-Mart\'{\i}nez}},\
  }\href {\doibase 10.1103/PhysRevD.90.064003} {\bibfield  {journal} {\bibinfo
  {journal} {Phys. Rev. D}\ }\textbf {\bibinfo {volume} {90}},\ \bibinfo
  {pages} {064003} (\bibinfo {year} {2014})}\BibitemShut {NoStop}%
\bibitem [{\citenamefont {Valentini}(1991)}]{Valentini1991}%
  \BibitemOpen
  \bibfield  {author} {\bibinfo {author} {\bibfnamefont {A.}~\bibnamefont
  {Valentini}},\ }\href {\doibase
  http://dx.doi.org/10.1016/0375-9601(91)90952-5} {\bibfield  {journal}
  {\bibinfo  {journal} {Phys. Lett. A}\ }\textbf {\bibinfo {volume} {153}},\
  \bibinfo {pages} {321 } (\bibinfo {year} {1991})}\BibitemShut {NoStop}%
\bibitem [{\citenamefont {Reznik}(2003)}]{Reznik2003}%
  \BibitemOpen
  \bibfield  {author} {\bibinfo {author} {\bibfnamefont {B.}~\bibnamefont
  {Reznik}},\ }\href {\doibase 10.1023/A:1022875910744} {\bibfield  {journal}
  {\bibinfo  {journal} {Found. Phys.}\ }\textbf {\bibinfo {volume} {33}},\
  \bibinfo {pages} {167} (\bibinfo {year} {2003})}\BibitemShut {NoStop}%
\bibitem [{\citenamefont {Reznik}\ \emph {et~al.}(2005)\citenamefont {Reznik},
  \citenamefont {Retzker},\ and\ \citenamefont {Silman}}]{Reznik1}%
  \BibitemOpen
  \bibfield  {author} {\bibinfo {author} {\bibfnamefont {B.}~\bibnamefont
  {Reznik}}, \bibinfo {author} {\bibfnamefont {A.}~\bibnamefont {Retzker}}, \
  and\ \bibinfo {author} {\bibfnamefont {J.}~\bibnamefont {Silman}},\ }\href
  {http://link.aps.org/abstract/PRA/v71/e042104} {\bibfield  {journal}
  {\bibinfo  {journal} {Phys. Rev. A}\ }\textbf {\bibinfo {volume} {71}},\
  \bibinfo {eid} {042104} (\bibinfo {year} {2005})}\BibitemShut {NoStop}%
\bibitem [{\citenamefont {Pozas-Kerstjens}\ and\ \citenamefont
  {Mart\'{i}n-Mart\'{i}nez}(2015)}]{Pozas-Kerstjens:2015}%
  \BibitemOpen
  \bibfield  {author} {\bibinfo {author} {\bibfnamefont {A.}~\bibnamefont
  {Pozas-Kerstjens}}\ and\ \bibinfo {author} {\bibfnamefont {E.}~\bibnamefont
  {Mart\'{i}n-Mart\'{i}nez}},\ }\href {\doibase 10.1103/PhysRevD.92.064042}
  {\bibfield  {journal} {\bibinfo  {journal} {Phys. Rev. D}\ }\textbf {\bibinfo
  {volume} {92}},\ \bibinfo {pages} {064042} (\bibinfo {year}
  {2015})}\BibitemShut {NoStop}%
\bibitem [{\citenamefont {Pozas-Kerstjens}\ and\ \citenamefont
  {Mart\'{i}n-Mart\'{i}nez}(2016)}]{Pozas2016}%
  \BibitemOpen
  \bibfield  {author} {\bibinfo {author} {\bibfnamefont {A.}~\bibnamefont
  {Pozas-Kerstjens}}\ and\ \bibinfo {author} {\bibfnamefont {E.}~\bibnamefont
  {Mart\'{i}n-Mart\'{i}nez}},\ }\href {\doibase 10.1103/PhysRevD.94.064074}
  {\bibfield  {journal} {\bibinfo  {journal} {Phys. Rev. D}\ }\textbf {\bibinfo
  {volume} {94}},\ \bibinfo {pages} {064074} (\bibinfo {year}
  {2016})}\BibitemShut {NoStop}%
\bibitem [{\citenamefont {VerSteeg}\ and\ \citenamefont
  {Menicucci}(2009)}]{Nick}%
  \BibitemOpen
  \bibfield  {author} {\bibinfo {author} {\bibfnamefont {G.}~\bibnamefont
  {VerSteeg}}\ and\ \bibinfo {author} {\bibfnamefont {N.~C.}\ \bibnamefont
  {Menicucci}},\ }\href@noop {} {\bibfield  {journal} {\bibinfo  {journal}
  {Phys. Rev. D}\ }\textbf {\bibinfo {volume} {79}},\ \bibinfo {pages} {044027}
  (\bibinfo {year} {2009})}\BibitemShut {NoStop}%
\bibitem [{\citenamefont {Mart\'{\i}n-Mart\'{\i}nez}\ \emph
  {et~al.}(2016)\citenamefont {Mart\'{\i}n-Mart\'{\i}nez}, \citenamefont
  {Smith},\ and\ \citenamefont {Terno}}]{topology}%
  \BibitemOpen
  \bibfield  {author} {\bibinfo {author} {\bibfnamefont {E.}~\bibnamefont
  {Mart\'{\i}n-Mart\'{\i}nez}}, \bibinfo {author} {\bibfnamefont {A.~R.~H.}\
  \bibnamefont {Smith}}, \ and\ \bibinfo {author} {\bibfnamefont {D.~R.}\
  \bibnamefont {Terno}},\ }\href {\doibase 10.1103/PhysRevD.93.044001}
  {\bibfield  {journal} {\bibinfo  {journal} {Phys. Rev. D}\ }\textbf {\bibinfo
  {volume} {93}},\ \bibinfo {pages} {044001} (\bibinfo {year}
  {2016})}\BibitemShut {NoStop}%
\bibitem [{\citenamefont {Jonsson}\ \emph {et~al.}(2018)\citenamefont
  {Jonsson}, \citenamefont {Ried}, \citenamefont
  {Mart{\'{i}}n-Mart{\'{i}}nez},\ and\ \citenamefont {Kempf}}]{Katja}%
  \BibitemOpen
  \bibfield  {author} {\bibinfo {author} {\bibfnamefont {R.~H.}\ \bibnamefont
  {Jonsson}}, \bibinfo {author} {\bibfnamefont {K.}~\bibnamefont {Ried}},
  \bibinfo {author} {\bibfnamefont {E.}~\bibnamefont
  {Mart{\'{i}}n-Mart{\'{i}}nez}}, \ and\ \bibinfo {author} {\bibfnamefont
  {A.}~\bibnamefont {Kempf}},\ }\href {\doibase 10.1088/1751-8121/aae78a}
  {\bibfield  {journal} {\bibinfo  {journal} {J. Phys. A}\ }\textbf {\bibinfo
  {volume} {51}},\ \bibinfo {pages} {485301} (\bibinfo {year}
  {2018})}\BibitemShut {NoStop}%
\bibitem [{\citenamefont {Simidzija}\ \emph {et~al.}(2020)\citenamefont
  {Simidzija}, \citenamefont {Ahmadzadegan}, \citenamefont {Kempf},\ and\
  \citenamefont {Mart\'{i}n-Mart\'{i}nez}}]{Simidzija_2020}%
  \BibitemOpen
  \bibfield  {author} {\bibinfo {author} {\bibfnamefont {P.}~\bibnamefont
  {Simidzija}}, \bibinfo {author} {\bibfnamefont {A.}~\bibnamefont
  {Ahmadzadegan}}, \bibinfo {author} {\bibfnamefont {A.}~\bibnamefont {Kempf}},
  \ and\ \bibinfo {author} {\bibfnamefont {E.}~\bibnamefont
  {Mart\'{i}n-Mart\'{i}nez}},\ }\href {\doibase 10.1103/PhysRevD.101.036014}
  {\bibfield  {journal} {\bibinfo  {journal} {Phys. Rev. D}\ }\textbf {\bibinfo
  {volume} {101}},\ \bibinfo {pages} {036014} (\bibinfo {year}
  {2020})}\BibitemShut {NoStop}%
\bibitem [{\citenamefont {Faure}\ \emph {et~al.}(2020)\citenamefont {Faure},
  \citenamefont {Perche},\ and\ \citenamefont {Torres}}]{remi}%
  \BibitemOpen
  \bibfield  {author} {\bibinfo {author} {\bibfnamefont {R.}~\bibnamefont
  {Faure}}, \bibinfo {author} {\bibfnamefont {T.~R.}\ \bibnamefont {Perche}}, \
  and\ \bibinfo {author} {\bibfnamefont {B.~d. S.~L.}\ \bibnamefont {Torres}},\
  }\href {\doibase 10.1103/PhysRevD.101.125018} {\bibfield  {journal} {\bibinfo
   {journal} {Phys. Rev. D}\ }\textbf {\bibinfo {volume} {101}},\ \bibinfo
  {pages} {125018} (\bibinfo {year} {2020})}\BibitemShut {NoStop}%
\bibitem [{\citenamefont {Wald}(1984)}]{Wald1}%
  \BibitemOpen
  \bibfield  {author} {\bibinfo {author} {\bibfnamefont {R.~M.}\ \bibnamefont
  {Wald}},\ }\href@noop {} {\emph {\bibinfo {title} {General Relativity}}}\
  (\bibinfo  {publisher} {The University of Chicago Press},\ \bibinfo {year}
  {1984})\BibitemShut {NoStop}%
\bibitem [{\citenamefont {Weber}(1961)}]{weber}%
  \BibitemOpen
  \bibfield  {author} {\bibinfo {author} {\bibfnamefont {J.}~\bibnamefont
  {Weber}},\ }\href@noop {} {\bibfield  {journal} {\bibinfo  {journal} {Inc.,
  New York}\ } (\bibinfo {year} {1961})}\BibitemShut {NoStop}%
\bibitem [{\citenamefont {Peskin}\ and\ \citenamefont
  {Schroeder}(1995)}]{peskin}%
  \BibitemOpen
  \bibfield  {author} {\bibinfo {author} {\bibfnamefont {M.~E.}\ \bibnamefont
  {Peskin}}\ and\ \bibinfo {author} {\bibfnamefont {D.~V.}\ \bibnamefont
  {Schroeder}},\ }\href@noop {} {\emph {\bibinfo {title} {{An Introduction to
  quantum field theory}}}}\ (\bibinfo  {publisher} {Addison-Wesley},\ \bibinfo
  {address} {Reading, USA},\ \bibinfo {year} {1995})\BibitemShut {NoStop}%
\bibitem [{\citenamefont {Perche}\ and\ \citenamefont {Neuser}(2021)}]{jonas}%
  \BibitemOpen
  \bibfield  {author} {\bibinfo {author} {\bibfnamefont {T.~R.}\ \bibnamefont
  {Perche}}\ and\ \bibinfo {author} {\bibfnamefont {J.}~\bibnamefont
  {Neuser}},\ }\href@noop {} {\enquote {\bibinfo {title} {A wavefunction
  description for a localized quantum particle in curved spacetimes},}\ }
  (\bibinfo {year} {2021}),\ \Eprint {http://arxiv.org/abs/2012.08539}
  {arXiv:2012.08539 [gr-qc]} \BibitemShut {NoStop}%
\bibitem [{\citenamefont {Satz}(2007)}]{Satz_2007}%
  \BibitemOpen
  \bibfield  {author} {\bibinfo {author} {\bibfnamefont {A.}~\bibnamefont
  {Satz}},\ }\href {\doibase 10.1088/0264-9381/24/7/003} {\bibfield  {journal}
  {\bibinfo  {journal} {Class. Quantum Gravity}\ }\textbf {\bibinfo {volume}
  {24}},\ \bibinfo {pages} {1719} (\bibinfo {year} {2007})}\BibitemShut
  {NoStop}%
\bibitem [{\citenamefont {Louko}\ and\ \citenamefont
  {Satz}(2008)}]{LoukoCurvedSpacetimes}%
  \BibitemOpen
  \bibfield  {author} {\bibinfo {author} {\bibfnamefont {J.}~\bibnamefont
  {Louko}}\ and\ \bibinfo {author} {\bibfnamefont {A.}~\bibnamefont {Satz}},\
  }\href {\doibase 10.1088/0264-9381/25/5/055012} {\bibfield  {journal}
  {\bibinfo  {journal} {Class. Quantum Gravity}\ }\textbf {\bibinfo {volume}
  {25}},\ \bibinfo {pages} {055012} (\bibinfo {year} {2008})}\BibitemShut
  {NoStop}%
\bibitem [{\citenamefont {Tjoa}\ \emph {et~al.}(2021)\citenamefont {Tjoa},
  \citenamefont {L\'opez-Guti\'errez}, \citenamefont {Sachs},\ and\
  \citenamefont {Mart\'{\i}n-Mart\'{\i}nez}}]{erickson}%
  \BibitemOpen
  \bibfield  {author} {\bibinfo {author} {\bibfnamefont {E.}~\bibnamefont
  {Tjoa}}, \bibinfo {author} {\bibfnamefont {I.}~\bibnamefont
  {L\'opez-Guti\'errez}}, \bibinfo {author} {\bibfnamefont {A.}~\bibnamefont
  {Sachs}}, \ and\ \bibinfo {author} {\bibfnamefont {E.}~\bibnamefont
  {Mart\'{\i}n-Mart\'{\i}nez}},\ }\href {\doibase 10.1103/PhysRevD.103.125021}
  {\bibfield  {journal} {\bibinfo  {journal} {Phys. Rev. D}\ }\textbf {\bibinfo
  {volume} {103}},\ \bibinfo {pages} {125021} (\bibinfo {year}
  {2021})}\BibitemShut {NoStop}%
\end{thebibliography}%

\end{document}